# A New Qubits Mapping Mechanism for Multi-programming Quantum Computing

Xinglei Dou, Lei Liu
*Sys-Inventor Lab, SKLCA, ICT, CAS, Beijing, China*
*lei.liu@zoho.com; liulei2010@ict.ac.cn*
*Corresponding Author is Lei Liu (PI)*

*Abstract* - For a specific quantum chip, multi-programming helps to improve overall throughput and resource utilization. However, the previous solutions for mapping multiple programs onto a quantum chip often lead to resource under-utilization, high error rate and low fidelity. In this paper, we propose a new approach to map concurrent quantum programs. Our approach has three critical components. The first one is the Community Detection Assisted Partition (CDAP) algorithm, which partitions physical qubits for concurrent quantum programs by considering both physical typology and the error rates, avoiding the waste of robust resources. The second one is the X-SWAP scheme that enables inter-program SWAP operations to reduce the SWAP overheads. Finally, we propose a compilation task scheduling framework, which dynamically selects concurrent quantum programs to be executed based on estimated fidelity, increasing the throughput of the quantum computer. We evaluate our work on publicly available quantum computer IBMQ16 and a simulated quantum chip IBMQ20. Our work outperforms the previous solution on multi-programming in both fidelity and SWAP overheads by 12.0% and 11.1%, respectively.

## I. INTRODUCTION

Quantum computer is stepping into our view. Due to its potential in various critical applications, such as quantum machine learning [5,18], database search [11] and chemistry simulation [13,24], many companies, universities and institutes have joined the community to develop kinds of quantum computers. Google has released their quantum chip with 72 quantum bits (qubits) [14], which is with higher number of qubits than IBM (50 qubits) [15] and Intel (49 qubits) [12]. Recently, IBM announced its quantum chip with 53 quantum bits (qubits) accessible via the cloud [10]. However, the modern quantum chip belongs to Noisy Intermediate-Scale Quantum (NISQ) category [25] – the qubits and the links between them are not with the same reliability and are easily to be disturbed and therefore quantum computers are susceptible to errors. Through quantum computers can be made fault-tolerant by leveraging quantum error correction (QEC) codes, such codes are too expensive (20-100 physical qubits to form a fault-tolerant logical qubit) to be implemented on NISQ [9].

Cloud is everywhere in this age. In front of cloud computing, people like to put everything on the cloud, and quantum computers are no exception. With the growing access to quantum computers, it is necessary to increase the throughput of quantum computers that are with a limited number of qubits. Previous work [7] studies mapping multiple quantum programs on a specific quantum chip for improving the utilization. In this paper, we study multi-programming as a means to improve the throughput of the quantum computers. Although mapping multiple quantum programs onto a specific quantum chip improves the throughput, the activity of one program can negatively affect the reliability of a co-executing program because of (1) limited resources with high fidelity are scarce for quantum programs; (2) crosstalk noise caused by simultaneously executed quantum gates [21]; and (3) long SWAP paths. Previous studies [7] shows that running multiple quantum programs on a specific quantum chip incurs 12% reduction in fidelity on average.

The goal of this paper is to propose solutions that improve the throughput and utilization of NISQ machines while reducing the negative impacts on reliability associated with multi-programming NISQ computers. We find the previous qubit mapping policies have several shortcomings when used to map multiple quantum programs. (1) The existing mapping policies often divide a large area of robust on-chip qubits into many small-scale segments that cannot be mapped onto for any other programs. On average, over 20% of the robust qubits are wasted during the initial mapping. (2) When a specific quantum chip is partitioned into segments for mapping multiple quantum programs, post-compilation SWAP operations for each quantum program can be more, leading to an unpredictable impact for fidelity and reliability. For instance, several additional SWAPs can be incurred when two quantum programs with tens of CNOT gates are compiled together.

To this end, we design a new qubits mapping mechanism, which has two key features. First, it partitions the physical qubits for concurrent quantum programs with community detection techniques, avoiding the waste caused by the typology-unaware algorithm. It also provides a better initial-mapping, which reduces the SWAP overheads. Second, it uses the Inter-program SWAP operations to achieve qubit entangling, which reduces the overall SWAP overheads in multi-programming cases. Our approach works well in practice. The experimental results show that our approach outperforms previous solution [7] by 12.0% in fidelity and 11.1% in post-compilation gates reduction.

To sum up, we make the following contributions in this work. (1) We observe that the previous qubits allocation policies cannot make full utilization of resources in the cases where multiple quantum programs are running on a specific quantum chip at the same time. (2) We reveal that, rather than the independent execution cases, the number of SWAP operations increases significantly in the cases where multiple quantum programs are running together. (3) We propose a new qubits mapping approach (CDAP), which provides better mapping policies than the latest work for co-located quantum programs. Further, we design X-SWAP, the first inter-program SWAP mechanism, which can significantly reduce the SWAP overheads. (4) We propose a compilation task-scheduling framework, which selects the optimal workload combinations for multi-programming, maximizing the fidelity and resource utilization.

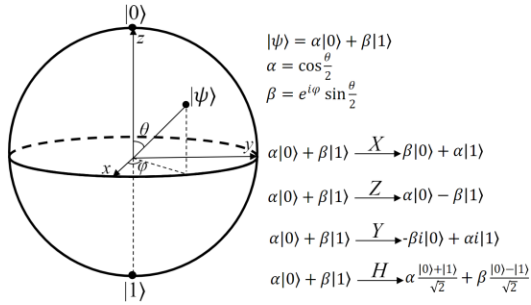

Fig. 1. Bloch sphere representation of a qubit.

## II. BACKGROUND

In this section, we introduce some concepts about quantum computers and quantum computing. We also introduce the relevant background of quantum cloud services to illustrate why multi-programming in NISQ computers is necessary.

### A. Quantum mechanism

Quantum computing can solve conventionally hard problems by leveraging quantum mechanisms [27]. The foundation of quantum computing lies on quantum bits (qubits). As shown in figure 1, The Bloch sphere is used to represent the state of a single qubit. The north and south poles of the sphere represent the $|0\rangle$ and $|1\rangle$ states of the qubit, respectively, corresponding to the 0 and 1 in classical bits. The state of a qubit can be the linear combinations of $|0\rangle$ and $|1\rangle$ as $|\psi\rangle = \alpha |0\rangle + \beta |1\rangle$, where $\alpha, \beta$ are complex numbers with $|\alpha|^2 + |\beta|^2 = 1$. The state of the qubits can be manipulated with Single-qubit gates, e.g., $H$, $X$, $Z$, etc. When the qubit is measured, the state of the qubit collapse to $|0\rangle$ and $|1\rangle$ with the possibility of $|\alpha|^2$ and $|\beta|^2$, respectively. Two or more qubits can be entangled with two-qubit gates - Control-NOT (CNOT) gates. A CNOT gate flips the state of the target qubits when the control qubit is in the state $|1\rangle$. Likewise, the state of a two-qubit system is represented by: $|\psi\rangle = \alpha_{00} |00\rangle + \alpha_{01} |01\rangle + \alpha_{10} |10\rangle + \alpha_{11} |11\rangle$. A quantum gate manipulating three or more qubits can be decomposed with single- and two-qubit gates [4].

### B. Quantum computers

There are several competing technologies for the physical implementation of quantum computers. Among them, the most prevalent two are ion-trap [8] and superconducting quantum circuits [16]. IBM quantum processors are superconducting quantum chips with Josephson-junction-based transmon qubits [16] and microwave-tunable two-qubit gates [26]. Different from the all-to-all connectivity of qubits in ion-trap quantum computers, the physical qubits in IBM quantum processors only have connections to the neighboring qubits. For example, Figure 2 shows the architecture of IBM Q16 Melbourne and previously available IBM Q20 Tokyo (referred to as IBMQ16 and IBMQ20 Errors in quantum computers hereafter) quantum chip. IBMQ20 is provided in IBM qiskit compiler [2] as a fake backend.

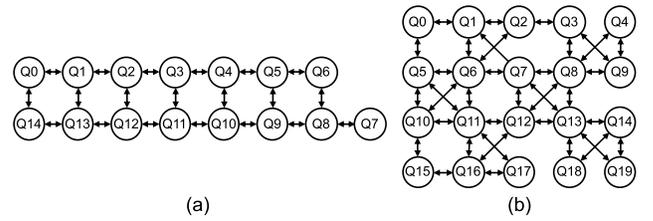

Fig. 2. (a) IBM Q16 Melbourne architecture. (b) Previous available IBM Q20 Tokyo architecture.

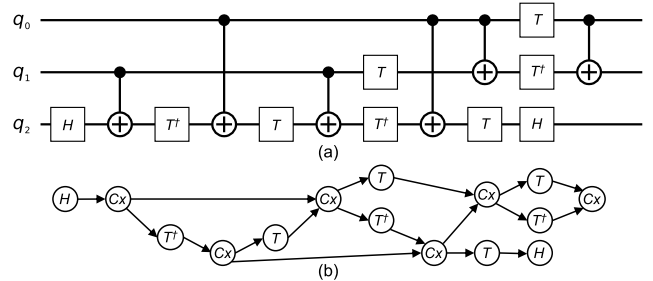

Fig. 3. The Quantum circuit and DAG of decomposed Toffoli gate.

### C. Errors in quantum computers

NISQ computers face reliability challenges. As the physical qubits are fragile and susceptible to interference, the following kinds of errors may occur in quantum programs running on quantum computers: (1) Coherence errors caused by short qubit state retention time. (2) Operational errors caused by error-prone quantum gates. (3) Readout errors caused by measurement operations. In practice, the error rates and the coherence time are all reported in IBM backends calibration data [1]. The error rates in IBM backends are variational, both spatially and temporally. Variation-aware mapping policies tend to map quantum programs onto regions with the lowest error rate.

### D. Quantum programs

A quantum program consists of a sequence of quantum gates, which can be represented by a quantum circuit. For example, figure 3-(a) shows the quantum circuit of the decomposed Toffoli gate, where each horizontal line represents a qubit, each block and vertical line represents a single- and two-qubit gate, respectively. (b) shows the Directed Acyclic Graph (DAG) of the circuit. A quantum gate is logically executable when it has no unexecuted predecessors in the DAG. A CNOT operation cannot be executed unless the two logical qubits it associated with are mapped physically adjacent. Generally speaking, it takes two steps for a quantum program compiler to solve the mapping problem: (1) Initial mapping generation. The compiler maps each program qubit onto a physical qubit. (2) Mapping transition. The compiler satisfies all two-qubit constraints by inserting SWAPs (A SWAP can be decomposed to three CNOT gates [19]) to the quantum circuit, so that any two-qubit gates in the quantum program can be executed physically.

### E. Quantum cloud services

It takes huge cost to maintain a quantum computer, but vendors provide easy-to-use interfaces, which enables users to have access to quantum computers through cloud services conveniently. Take IBM Q Experience [1] provided by IBM as an example, the cloud service provides multiple backends, each of which corresponds to a quantum computer. Quantum tasks can be constructed and executed on the target quantum computer with the help of qiskit framework [2]. However, due to the scarcity of quantum computing resources and the growing need for quantum computer access, contention for accessing quantum computers increases. For example, for backend IBM Q Vigo, we observe an average of more than 120 queued jobs in a day. The throughput and utilization of quantum computers need to be improved to help shorten the average queuing time and handle more tasks.

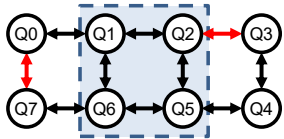

Fig. 4. A case for qubits mapping. Highlighted links with red color represent links with high error rates.

## III. MOTIVATION

Quantum computers are susceptible to errors (Noisy Intermediate Scale Quantum, NISQ), as the state of the qubit can only keep in a short time (30-100 μs) [1]. In reality, the error rates are variational across all of the qubits and links in NISQ computers; not all of the qubits are with the similar (reliable) states, and not all of the links between qubits are with the same error rates.

To map a specific quantum program, previous noise-aware mapping techniques for the single quantum program [20,23,29] usually employ greedy or heuristic approaches to discover the mapping policies that have the most reliable qubits and links. For example, in Figure 4, $Q_1$, $Q_2$, $Q_5$ and $Q_6$ are mapped as they are with better performance, and the links between them are with lower error rates.

However, for a chip consists of tens of the qubits, e.g., 72 qubits, if a mapping policy is not carefully designed, the qubits - especially these robust ones - will likely be wasted, and therefore the overall throughput of the quantum computer will be low. In this era, the resource utilization problem is equal to that of banquet in cloud environments; we should try to avoid those wasting any resources and make a higher utilization of resources on quantum chips. Thus, for a chip with tens of qubits (or more), people try to map multiple quantum programs on it. Yet, mapping multiple quantum programs can be a challenging job. The co-located quantum programs exhibit different characteristics, including quantum circuit depth, number of qubits, number of two-qubit gates, etc. The reliable resources are limited and the fairness of allocation for concurrent quantum programs needs to be guaranteed. When multiple quantum programs are mapped onto a specific quantum chip, a significant reduction in fidelity could be incurred [7]. Therefore, the mapping policy must be designed and optimized to achieve a better reliability – lower error rates – for concurrently running quantum programs.

How to design the scheme for mapping multiple quantum programs on a specific chip simultaneously? – The existing multi-programming technique [7] supports to co-locate 2 quantum programs. However, if more than two quantum programs are running together, the existing approaches can hardly have an ideal solution for **co-locating them with lower error rates and high reliability.** Some previous studies propose solutions for mapping single quantum program [17,29]. Nevertheless, the principle for mapping multiple programs is quite different from the scheme that maps a single program. The previous work mainly relies on heuristic policies and greedy algorithms. However, in terms of multiple co-running quantum programs, these studies can use improvements. In this paper, we summarize the challenges and introduce our insights as follows.

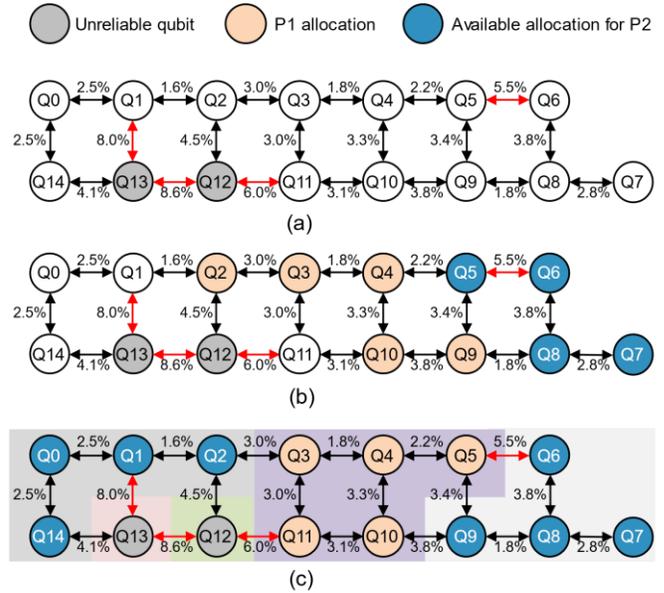

Fig. 5. A case for qubits mapping. Highlighted links with red color represent links with high error rates.

### A. Resource under-utilization

Existing mapping policies face the challenge of resource under-utilization. The available qubits are subject to be divided into smaller scale segments. Some of them can support to map a quantum program, but some of them cannot due to the high error rates and weak links. Hence, when mapping multiple quantum programs, some of the programs might have weak qubits, leading to unreliable results and increasing error rates for co-running programs. The latest study [7] proposes how to map multiple quantum programs onto a specific quantum chip. Figure 5 shows an example of the previous approach on how to map quantum programs $P_1$ (5 qubits) and $P_2$ (4 qubits) on IBMQ16. The unreliable qubits and weak links are highlighted in Figure 5-(a). Figure 5-(b) illustrates the results of using the mapping approach (FRP) in [7]. For $P_1$, FRP tries to have the most reliable qubits and reliable links started from a reliable root by using a greedy policy according to *utility* (denoted by the value of the number of links belong to a specific qubit / the sum of error rates of CNOT operations of the qubit). As a result, $Q_2 - Q_4$, $Q_9$, and $Q_{10}$ are involved. However, after the allocation for $P_1$, the FRP cannot avoid using the unstable/unreliable qubits and links; thus $Q_5 - Q_8$ are allocated for $P_2$. Some of the reliable quantum bits and links are wasted, e.g., $Q_{11}$, $Q_{14}$. Please note that if the simple greedy algorithm is employed, the results can be even worse. In fact, there exist better solutions for mapping $P_1$ and $P_2$. Figure 5-(c) illustrates a better solution, in which $P_2$ avoids the unreliable link, and $P_1$'s mapping still has reliable qubits and links. Apparently, the existing qubit mapping scheme can incur qubits under-utilization, wasting resources.

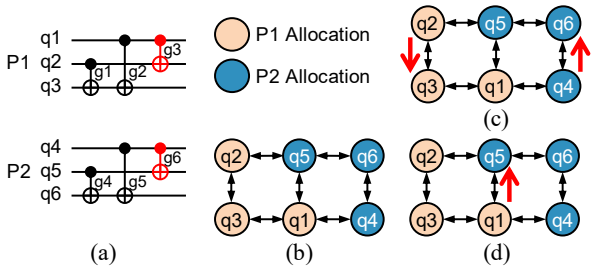

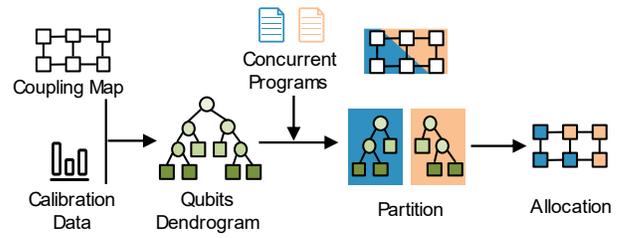

Fig. 6. A case for qubits mapping. Highlighted links with red color represent links with high error rates.

Fig. 7. Qubits mapping by using CDAP algorithm in a nutshell.

*B. More SWAP operations - high SWAP overheads*

When multiple quantum programs are running concurrently, the number of SWAP operations involved for a specific program can be higher than that in the single running cases. The previous multi-programming policy [7] compiles quantum programs one by one; the next program must be mapped to the unused qubits. And, the SWAP operations for one program cannot involve the qubits mapped to other programs. Roughly speaking, the SWAP path for a specific quantum program can involve several qubits, among which the SWAP operations can only be conducted between the neighboring qubits. If any qubit on the shortest SWAP path is occupied by other programs, the whole SWAP operations have to suffer higher overheads, i.e., involving more SWAP operations across more qubits. Known to us, more SWAP operations bring higher error rates [29], therefore the overall fidelity is negatively impacted as a result.

Figure 6 shows a case where two quantum programs are mapped on a specific quantum chip. Figure 6-(a) shows the circuits of two quantum programs ($P_1$ and $P_2$), and (b) illustrates how they are mapped on a quantum chip with 6 physical bits. According to the quantum circuits, for $P_1$, the $g_1$ and $g_2$ can be executed directly – without any SWAP operations involved. However, the highlighted $g_3$ cannot be executed directly unless a SWAP operation for $q_2$ and $q_3$ is executed firstly. Same thing happens for $P_2$. A SWAP operation between $q_4$ and $q_6$ should be involved before $g_6$ can be executed. To sum up, for such a simple qubits mapping case, two SWAP operations are needed. Figure 6-(c) shows the SWAP operations.

However, if multiple programs could be compiled together, an inter-program SWAP operation can be helpful. Figure 6-(d) shows a new policy that enables inter-program SWAP. Illustrated in (d), only one inter-program SWAP $\{q_1, q_5\}$ is needed, and thus all of the quantum gates can be executed without other overheads. As mentioned before, SWAP operation leads to high error rate and incurs unreliable problems (each SWAP operation has three CNOT gates, and more CNOT gates will also bring high error rates [29]). Less SWAP operations can benefit the overall performance for a specific quantum computer and also the computing results.

*C. Related work*

Recent works towards mapping quantum programs onto quantum chips mainly relies on heuristic search schemes. IBM's compilation framework [2] implements noisy adaptive mapping [20] and Stochastic SWAP. [31] generates mapping transition by inserting SWAP layers into adjacent data independent layers. [29] enhances [31] by introducing noise variation awareness. SABRE [17] brings exponential speedup in the search complexity by reducing the search space. QURE [3] finds reliable partitions by searching isomorphic sub-graphs. These approaches focus on mapping problem for only one quantum program. However, new policies need to be proposed for mapping multiple quantum programs and ensure the fairness and reliability at the same time.

For multi-programming, [7] proposes FRP algorithm to assign reliable regions for each quantum program. It applies enhanced SABRE [17] with noise variation awareness to generate the mapping transition. However, the approach cannot make a good use of the robust qubits on chip, the utilization and the throughput can be further improved.

In our work, we design a new policy to generate a better initial mapping for multi-programing quantum workloads and design a new SWAP approach. Our work makes better use of the qubit resources and significantly improve the quantum computers' throughput.

## IV. THE ART OF OUR DESIGN

*A. The design of a new qubits mapping policy*

The initial mapping is critical for a specific quantum program. An excellent initial mapping helps to reduce the SWAP overheads and can also make full use of the robust qubits and links on the quantum chips. In terms of the concurrent quantum programs, an excellent initial mapping not only reduces the SWAP cost during mapping, but also reduces the interference between multiple concurrent programs and improve the overall fidelity.

We have the following observations about quantum chip and multi-programming. (1) Concurrent quantum programs are independent of each other. (2) The robust qubits and links on a specific quantum chip are limited. (3) Some qubits on the quantum chip have more connections to their surroundings, e.g., as shown in Figure 2-b, $Q_{11}$ has links to the 6 adjacent physical qubits, while $Q_0$ has links to only 2 qubits. (4) The qubits needed for a single program should to be closely allocated. The allocations for qubits belongs to different quantum programs should avoid mutual interference, fairly leveraging of the robust links.

To this end, we propose a new technology - Community Detection Assisted Partitioning (CDAP) - to construct a hierarchy tree consisted of qubits for searching the robust qubits that are tightly connected for initial allocation. Figure 7 shows CDAP. CDAP creates a hierarchy tree according to the coupling

```
Algorithm 1: Hierarchy Tree Construction.
  Input: Chip Coupling Graph G, Calibration data C
  Output: The root of the dendrogram root
1 Communities ← [empty];
2 Create a leaf node for every qubit in Communities;
3 while Communities.length > 1 do
4    Find a two-element combination (A, B) of
     Communities with the max value of F(A, B);
5    New_Node ← union(A, B);
6    Set A and B as New_Node's left subtree and right
     subtree, respectively;
7    Remove A and B from Communities and append
     New_Node to Communities;
8 end
9 root ← Communities[0];
10 Return root.
```

```
Algorithm 2: Qubits Partitioning.
  Input: The hierarchy tree root, Quantum programs circuits
  Output: Partition partition
1 Rank circuits in descending order of CNOT density;
2 for circuit in circuits do
3    candidates ← set();
4    Find the all leaf nodes leaves of root;
5    for leaf in leaves do
6       while leaf is not None do
7          if qubits in leaf is enough for circuit then
8             candidates.add(leaf);
9             Break;
10         else
11            leaf ← leaf.parent;
12         end
13      end
14   end
15   if candidates is empty then
16      Fail and revert back to independent execution;
17   end
18   Find the best candidate from candidates based on
     average fidelity;
19   partition.append(candidate);
20   Remove qubits of candidate from all nodes in the
     hierarchy tree;
21   if candidates.sibling has no path to other nodes then
22      Remove qubits of candidates.sibling from its
        parent nodes;
23      candidates.sibling.parent ← None;
24   end
25 end
26 Return partition.
```

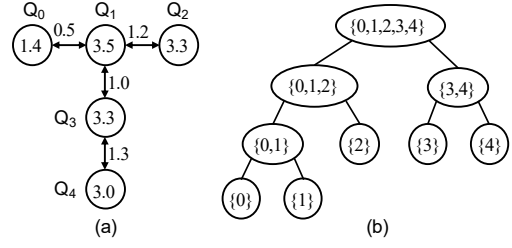

Fig. 8. (a) Calibration data of IBM Q London. The value in a node represents the readout error rate (in %) of the qubit, and the value on a link means the error rate (in %) of the CNOT operation. (b) The dendrogram generated by Algorithm 1. The values in nodes are the index of the physical qubits.

map and calibration data obtained from IBMQ API [1]. In the dendrogram of the hierarchy tree, a leaf node denotes a specific physical qubit, a circle node represents the union of its sub nodes. CDAP then iterates the hierarchy tree through bottom to top to find available regions for initial allocation. Finally, the quantum circuits are allocated by greedy policy to the corresponding regions. We show the details as below.

*1) Hierarchy tree construction:* The hierarchy tree is a profile of a quantum chip, which helps to locate reliable resources on the quantum computer. Algorithm 1 shows the hierarchy tree construction algorithm, which is based on FN community detection algorithm [22]. The algorithm clusters the physical qubits into communities. Qubits in a community have reliable and close interconnections. In contrast, the links between communities have relatively low fidelity.

When the algorithm starts, each physical qubit is actually a community, and is a leaf node in the dendrogram. The algorithm keeps merging two communities that can maximize the reward function $F$ until there is only one community containing all qubits. Each community during the process corresponds to a node in the hierarchy tree and is a candidate region for qubit allocation. The reward function $F$ is defined as the benefit of merging two communities:

$$F = Q_{merged} - Q_{origin} + \omega EV,$$

in which Q is the modularity of a partition (i.e., $Q = \sum_i (e_{ii} - a_i^2)$ [22], where $e_{ii}$ is the fraction of within-group edges in group $i$, and $a_i$ is the fraction of edges associated with vertices in group $i$). A higher $Q$ indicates a more proper partition. $Q_{origin}$ and $Q_{merged}$ denotes the modularity of the original partition and the new partition after merging the two communities, respectively. $E$ denotes the average fidelity (i.e., 1 minus the error rate of the operation) of CNOTs on the between-group edges of the two communities, and $V$ denotes the average fidelity of readout operations on the qubits connecting the two communities. The reward function $F$ enables CDAP to take both physical topology and error rate into account when performing partition. The $\omega$ is a weight parameter. For a specific quantum chip, we can change the value of $\omega$ to change the weight of physical typology and the error rate. If $\omega=0$, CDAP partitions completely based on physical typology. Noise-awareness is introduced as $\omega$ increases. If $\omega$ keeps increasing, the weight of the error rate will far exceed the weight of the physical typology, resulting in the degradation of CDAP into a greedy algorithm that is mainly based on error rate.

The hierarchy tree has several features: (1) Every node in the hierarchy tree is a candidate for initial allocation. (2) The physical qubits in a node are tightly-connected, which reduces the SWAP costs. (3) The qubits with a low readout error rate and robust links are preferentially merged. Thus, the more robust the qubit set is, the higher the node depth will be. The hierarchy tree helps to locate the robust resources on quantum computers, thus providing better initial mapping for quantum programs.

Further, we explain why the hierarchy tree helps selecting the initial allocation by an example in Figure 8. (i) $Q_0$ and $Q_1$ are firstly merged due to the link between them is with the lowest error rate. (ii) Then, $Q_2$ instead of $Q_3$ is added into the community {0,1}, though the link $Q_1$-$Q_3$ has lower error rates than $Q_1$-$Q_2$. This is because the algorithm tends to merged interconnected nodes into one community, avoiding the waste of robust physical qubits. Likewise, $Q_3$ and $Q_4$ are merged. (iii) Finally, all qubits are merged as the root of the hierarchy tree as illustrated in Figure 8-(b). The algorithm considers not only the robustness but also the physical topology, avoiding the waste of robust resources caused by the typology-unaware greedy algorithm, and supports more quantum programs to be mapped on a specific quantum chip.

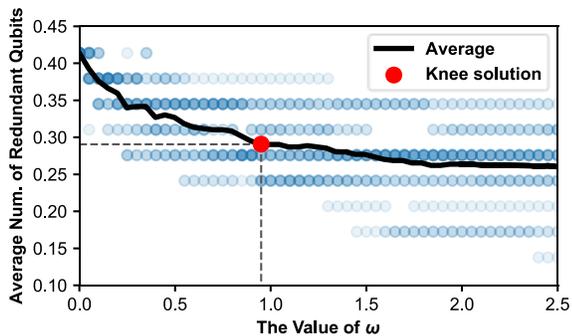

Fig. 9. The average number of redundant qubits in the hierarchy tree varies with $\omega$. A blue dot with a darker color represents more cases are overlapping in this result.

As the calibration data does not change all the time [1], the hierarchy tree construction process only needs to be performed once for each calibration. The hierarchy tree can be stored in the IBMQ cloud server or stored by the user for multiple quantum program mapping tasks.

*2) Partition and allocation:* Algorithm 2 partitions the qubits into multiple regions for concurrent quantum programs according to the hierarchy tree. The algorithm prioritizes the quantum programs with a higher value of the *CNOT density*, which is defined as (the number of CNOT instructions) / (the number of qubits in the quantum program). For each quantum program, the algorithm searches the hierarchy tree from bottom to top to find available candidate set of physical qubits. Then, the candidate with the lowest average error rate is assigned to the program for initial allocation. Finally, all of the quantum programs are allocated to the assigned regions with Greatest Weighted Edge First strategy [2].

*3) Discussion:* CDAP ensures that each community it forms has a tightly interconnected and robust set of physical qubits. Thus, the physical qubits are assigned as a group, which results in the number of physical qubits assigned to a quantum program can be larger than the number of qubits that a program really needs. For example, if a 4-qubit quantum program is mapped on the quantum chip in Figure 8, the only available region is the root of the hierarchy tree, i.e., {0,1,2,3,4}, leaving one redundant qubit unused. In order to avoid waste, we label the unused redundant qubits and they can be added to the adjacent communities.

For a node in a hierarchy tree, we define the term maximum redundant qubits, which refers to the maximum possible number of unused qubits when a quantum program is allocated to the community. The number of maximum redundant qubits of a node is computed as node.n_qubits-(1+max(node.left.n_qubits, node.right.n_qubits)). We observe the increase of $\omega$ in the reward function leads to the degradation of the hierarchy tree, i.e., in each merge process of the hierarchy tree, only one leaf node containing one qubit is added to the new community. The number of maximum redundant qubits of the new community is 0 in this case. Thus, the increase of $\omega$ leads to the reduction in average redundant qubits. We collected the calibration data of IBMQ16 for 21 days. The $\omega$ varies from 0 to 2.5 every day, and the average number of the redundant qubits in the hierarchy tree is recorded as shown in Figure 9. We take the knee solution, where the value of $\omega$ is 0.95. In this case, CDAP take both physical typology and the error rate into account, and the average redundant quantum number of the hierarchical tree is 0.29.

*B. The design of inter-program SWAP*

Multi-programming poses new challenges for mapping transition. We believe the mapping transition policies should make better use of possible SWAPs (both inter and intra-program), and minimize the circuit transformation overheads. In this paper, we design the X-SWAP, which includes both of the inter and intra-program SWAP operations, and provides the best SWAP solution by considering the SWAP possibilities across all of the qubits. In practice, the inter-program SWAP operation is enabled when two quantum programs are close to each other, and the cost of inter-program SWAP is less than the cost in the intra-program cases. The main advantage of X-SWAP over baseline is its expanded heuristic search space, which covers more qubits, thus enabling more possible SWAP operations. The below section shows the details.

*1) Why inter-program SWAP performs better:* In our study, we find there are two reasons behind this. (1) An inter-program SWAP can replace two or more intra-program SWAP operations in practice. As demonstrated in Figure 6-(c)/(d), the intra-program SWAPs, i.e., {$q_2$,$q_3$}, {$q_4$,$q_6$}, can be replaced by one inter-program SWAP across $q_5$ and $q_1$. Obviously, the inter-program SWAP incurs relatively lower overheads for the same goal. (2) Inter-program SWAPs take shortcuts. For instance, as shown in Figure 10-(e), an inter-program SWAP {$q_1$,$q_9$} takes only one step to satisfy CNOT $q_1$, $q_5$. However, in contrast, to achieve the same goal, previous intra-program scheme has to introduce three SWAPs, i.e., {$q_1$,$q_2$}, {$q_1$,$q_3$}, {$q_1$,$q_4$}. Briefly, in practice, enabling inter-program SWAPs can result in fewer SWAP operations in the cases where multiple quantum programs are mapped as neighbors on a specific quantum chip, therefore reducing the SWAP overheads and benefiting the overall fidelity.

*2) How does X-SWAP scheme work:* SWAP-based heuristic search scheme is employed in previous work for mapping transition [17]. We use it as the baseline. Instead of generating a scheduling for each quantum program separately and then merge them, we are the first to design an approach (in algorithm 3) for generating the global schedule for all of the programs simultaneously. The design details are shown as follows.

**Design of the heuristic search space.** For a quantum programs $P_i$, the DAG $DAG_i$ represents the data dependency of $P_i$; the

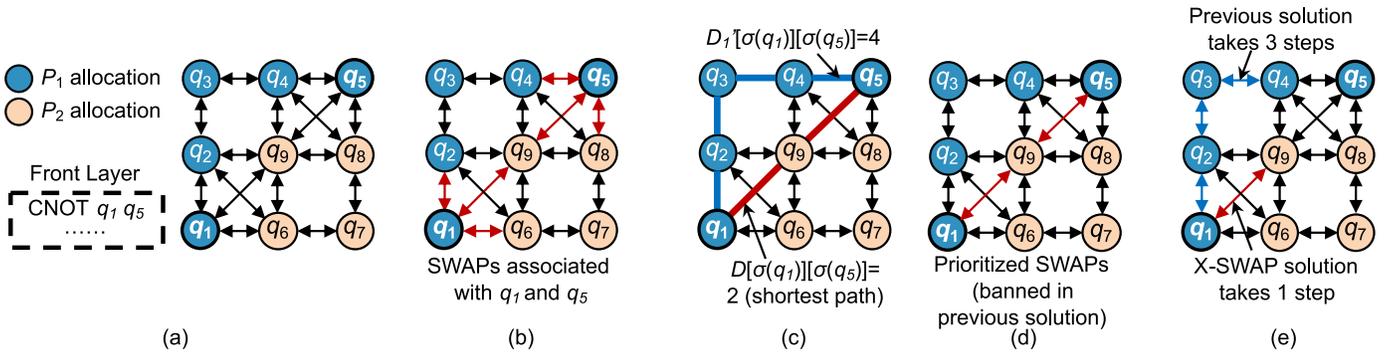

Fig. 10. (a) $P_1$ and $P_2$ are mapped on a quantum chip with 9 qubits. The next gate to be solved is CNOT that involves $q_1$ and $q_5$. (b) SWAPs associated with $q_1$ and $q_5$ are appended to SWAP candidates. (c) The gain of Inter-program SWAP is 2, $q_1$-$q_9$-$q_5$ is the shortest path. (d) SWAPs on the shortest path are prioritized. (e) X-SWAP scheme takes shortcuts to satisfy the constraint of CNOT $q_1$, $q_5$.

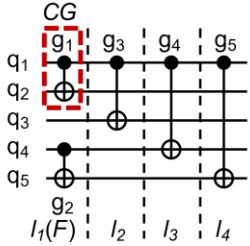

Fig. 11. A case for critical gates. The gates are in four layers. In the Front Layer $l_1$, $g_1$ is critical gate and $g_2$ is not. If $g_1$ is solved firstly, the data dependency of $g_3$ is resolved and $g_3$ is appended into the Front Layer. If $g_2$ is solved firstly, the Front Layer won't be updated.

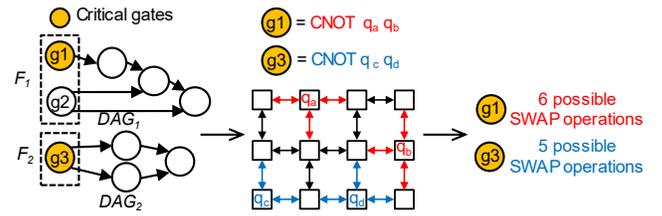

Fig. 12. Example of the mapping-transition procedure.

Front Layer $F_i$ denotes the set of all CNOT gates without unexecuted predecessors in $DAG_i$; and Critical Gates $CG_i$ denotes the CNOT gates in $F_i$ that have successors on the second data dependency layer as shown in Figure 11.

For each program $P_i$, the algorithm removes hardware-compliant gates that can be executed directly in $F_i$ from the $DAG_i$. When there are no hardware-compliant gates, SWAPs need to be inserted to make hardware-incompliant gates executable. Among all of the hardware-incompliant gates, critical gates need to be prioritized for the purpose of updating the Front Layer and reduce the post-compilation circuit depth. Thus, we only look for the SWAPs associated with qubits in critical CNOT gates. Figure 12 shows an example, in which $g_1$ and $g_3$ are critical gates. All SWAPs associated with critical gates are SWAP candidates. The best SWAP among candidates is selected according to the heuristic cost function. The mapping is updated when the SWAP is inserted into the circuit. Some hardware-incompliant CNOTs may become executable because their constraints are resolved with the SWAP. The procedure repeats until all the constraints of CNOTs in DAG are satisfied.

**Design of the heuristic cost function.** The heuristic cost function helps to find the best SWAP from all of the SWAP candidates in the heuristic search space. We show how to design a heuristic cost function which takes advantages of the characteristics in multi-programming to find reliable SWAPs.

The concept *locality* for mapping a quantum program is critical, indicating the mapping policy should keep qubits belong to a specific program close to each other. Otherwise, high SWAP overheads will occur between two qubits mapped far away from each other but are required for a CNOT operation. Nearest Neighbor Cost (NNC) function is used in SABRE [17] to estimate the SWAP cost to make hardware-incompliant gates executable. We also use the NNC function $H$ as a component in our heuristic function.

We also prioritize the SWAPs on the shortest path resolving the constraint of a CNOT gate with inter-program SWAPs. Given the coupling map of a quantum chip and the allocation of the quantum programs, we define the term distance matrix $D$, in which each cell denotes the shortest path between any two physical qubits on the quantum chip. For each program $P_i$, we define $D_i{'}$ as the shortest path matrix for unmapped physical qubits and the physical qubits on which $P_i$ is mapped. In essence, $D$ represents the shortest path matrix to perform mapping transition for concurrent quantum programs while allowing inter-program SWAPs, $D_i{'}$ represents the shortest path matrix to perform mapping transition for $P_i$ independently. For a two-qubit gate $g$, we denote the two logical qubits involved in $g$ as $g.q_1$ and $g.q_2$. We define the physical qubit on which a logical qubit $q$ is mapped as $\sigma(q)$. The shortest distance of two qubits involved in a two-qubit gate minus 1 is the minimum number of SWAPs required to satisfy its constraint.

In our design, if $D_i{'}[\sigma(g.q_1)][\sigma(g.q_2)]$ is greater than $D[\sigma(g.q_1)][\sigma(g.q_2)]$ for a two-qubit gate $g$ in a specific program $P_i$, it means that inter-program SWAPs outperforms intra-program SWAPs when satisfying the constraint of $g$. X-SWAP scheme should enable inter-program SWAPs in this case, which is beneficial to reduce the mapping transition cost of the quantum programs and brings lower overheads.

For example, in Figure 10, (a) shows two quantum programs are co-located on a chip. SWAP candidates associated with qubits in Front Layer are highlighted in (b). In (c) there is $D[\sigma(g.q_1)][\sigma(g.q_2)]=2$ and $D_i{'}[\sigma(g.q_1)][\sigma(g.q_2)]=4$, as it takes either 1 inter-program SWAP or 3 intra-program SWAPs to solve the constraint of CNOT $q_1$, $q_5$. We define the number of

**Algorithm 3:** X-SWAP framework.

**Input:** Coupling graph $G$, Quantum programs $circuits$, Initial mapping $mapping$
**Output:** Final schedule $schedule$

1. Generate DAG $DAG_i$ for each quantum program in $circuits$;
2. Generate Front Layer $F_i$ for each DAG;
3. **while** not all gates are resolved **do**
4.    **if** hardware-compliant gates exist **then**
5.       Append hardware-compliant gates to $schedule$;
6.       Remove hardware-compliant gates from the DAG and update Front Layer;
7.    **else**
8.       **for** $F_i$ in $F$ **do**
9.          Append gates in $F_i$ that have subsequent quantum gate operations on the second layer to $gates\_to\_resolve_i$;
10.         $SWAP\_candidate\_list_i \leftarrow$ Obtain_SWAPs($gates\_to\_resolve_i, G$);
11.       **end**
12.       Find a SWAP from all $SWAP\_candidate\_list_i$ with the max value of score(SWAP);
13.       Append the SWAP to $schedule$ and update $mapping$;
14.    **end**
15. **end**
16. Return $schedule$.

---

**Algorithm 4:** Task Scheduling.

**Input:** List of incoming jobs $jobs$, Hierarchy tree $root$

1. **while** $jobs$ is not empty **do**
2.    $cur\_job\_set=[jobs[0]]$;
3.    $idx \leftarrow 1$;
4.    **while** $idx < jobs.length$ and $idx < N$ and $cur\_job\_set.length < MAX\_COLOCATE\_NUM$ **do**
5.       $idx \leftarrow idx + 1$;
6.       $cur\_job\_set$.append($jobs[idx]$);
7.       **for** $job$ in $cur\_job\_set$ **do**
8.          $job.ind\_EPST \leftarrow$ get_ind_EPST($job$);
9.          $job.co\_EPST \leftarrow$ get_co_EPST($cur\_job\_set, job$);
10.         $EPST\_violation \leftarrow 1-(job.co\_EPST / job.ind\_EPST)$;
11.         **if** $EPST\_violation < \varepsilon$ **then**
12.            continue;
13.         **else**
14.            $cur\_job\_set$.remove($jobs[idx]$);
15.            goto line 4;
16.         **end**
17.       **end**
18.    **end**
19.    Algorithm 3 is called to compile programs in $cur\_job\_set$;
20.    $cur\_job\_set$ is submitted to the backends for execution;
21.    remove all programs in $cur\_job\_set$ from $jobs$.
22. **end**

---

SWAPs saved by X-SWAP scheme to satisfy the constraint of $g$ as:

$$gain(g)=D[\sigma(g.q_1)][\sigma(g.q_2)]-D_i'[\sigma(g.q_1)][\sigma(g.q_2)]$$

and, we define the heuristic cost function as:

$$score(\text{SWAP}) = H(\text{SWAP}) + \sum_{F_i \in F} \frac{1}{|F_i|} \sum_{g \in F_i} gain(g)I(\text{SWAP}, g)$$

The gain is normalized by the size of the Front Layer, as the size of different Front Layers are variable. $I$ (SWAP, g) = 1 when SWAP is on the shortest path satisfying the constraint of $g$ with the X-SWAP scheme, else $I$ (SWAP, g) = 0. SWAPs on the shortest path are prioritized as shown in Figure 10-(d). A SWAP with the minimum value of score is the best SWAP in the candidates.

In order to make strategies proposed above practical, we detail the overall framework as shown in Algorithm 3.

*C. Scheduling compilation tasks*

Although there exist some mapping mechanisms for multiple quantum programs [7], selecting appropriate quantum program combinations for multi-programming is still a challenging problem. The combinations are now specified by the user, which may cause the following defects: (1) Quantum computing resources are underutilized. (2) Randomly selected combinations may lead to a significant reduction in fidelity. (3) The results verification mechanism must be introduced, which brings additional system modification overheads. We propose a design for compilation task scheduler to select appropriate program combinations for multi-programming.

This design focuses on selecting optimal quantum program combinations, maximizing the fidelity and resource utilization of the quantum computer. For each task in the scheduling queue, the scheduler checks whether there are other quantum programs in the queue that can bring acceptable fidelity reduction when they are co-located on the quantum chip with the current task. If so, they are mapped to the target quantum computer simultaneously for execution. Otherwise, the task will be executed independently.

The Estimated Probability of a Successful Trial (EPST) is proposed to estimate the fidelity of the execution of a quantum program on a specific quantum chip. Independent EPST means the maximum EPST that the programs can achieve. To compute independent EPST, Algorithm 2 is called for every single quantum program to assign a set of physical qubits for it. On the contrary, co-located EPST represents the EPST when the programs are co-located on a quantum chip. Algorithm 2 is called to generate a partition for the programs. Error rates of quantum gates are reported in the calibration data. The EPST of a quantum program on a set of specified physical qubits is defined as follows:

$$EPST=f_{2q}^{|CNOTs|}f_{1q}^{|1q\ gates|}f_{ro}^{|qubits|}$$

where $f_{2q}$, $f_{1q}$ and $f_{ro}$ denotes the average fidelity of CNOTs, 1-qubit gates and readout operations on the specified physical qubits respectively; and $|CNOTs|$, $|1q\ gates|$ and $|qubits|$ denotes the number of CNOT gates, one-qubit gates and qubits of the quantum program respectively. A higher EPST indicates the quantum program is mapped to a region with more robust resources, and a higher PST may be obtained in actual execution.

TABLE I. NISQ BENCHMARKS

| Type | ID | Benchmark | Qubits | Num. of CNOTs | Num. of gates |
|---|---|---|---|---|---|
| Micro-sized | 1 | bv_n3 | 3 | 2 | 8 |
| | 2 | bv_n4 | 4 | 3 | 11 |
| | 3 | peres_3 | 3 | 7 | 16 |
| | 4 | toffoli_3 | 3 | 6 | 15 |
| | 5 | fredkin_3 | 3 | 8 | 16 |
| Small-sized | 6 | 3_17_13 | 3 | 17 | 36 |
| | 7 | 4mod5-v1_22 | 5 | 11 | 21 |
| | 8 | mod5mils_65 | 5 | 16 | 35 |
| | 9 | alu-v0_27 | 5 | 17 | 36 |
| | 10 | decod24-v2_43 | 4 | 22 | 52 |

TABLE II. PST COMPARISON BETWEEN MULTIPLE STRATEGIES.

| Workloads | | Independent | | | Baseline | | | CDAP+X-SWAP | | | CDAP-only | | | X-SWAP-only | | |
|---|---|---|---|---|---|---|---|---|---|---|---|---|---|---|---|---|
| ID1 | ID2 | PST1 | PST2 | avg | PST1 | PST2 | avg | PST1 | PST2 | avg | PST1 | PST2 | avg | PST1 | PST2 | avg |
| 1 | 1 | 71.7 | 71.9 | 71.8 | 47.7 | 52.0 | 49.9 | 71.7 | 76.7 | 74.2 | 71.5 | 75.5 | 73.5 | 55.0 | 51.9 | 53.4 |
| 1 | 2 | 71.3 | 31.4 | 51.4 | 71.0 | 22.9 | 47.0 | 71.7 | 35.0 | 53.3 | 72.2 | 35.4 | 53.8 | 67.1 | 21.2 | 44.2 |
| 1 | 3 | 71.8 | 78.3 | 75.1 | 44.8 | 41.2 | 43.0 | 48.3 | 52.7 | 50.5 | 40.6 | 53.1 | 46.8 | 47.9 | 39.6 | 43.7 |
| 1 | 4 | 72.0 | 83.0 | 77.5 | 44.1 | 25.4 | 34.8 | 74.9 | 81.5 | 78.2 | 75.1 | 81.1 | 78.1 | 43.9 | 30.2 | 37.1 |
| 1 | 5 | 71.7 | 75.9 | 73.8 | 47.2 | 40.6 | 43.9 | 37.3 | 78.6 | 57.9 | 44.3 | 79.1 | 61.7 | 47.1 | 38.8 | 43.0 |
| avg | | | 69.9 | | | 43.7 | | | 62.8 | | | 62.8 | | | 44.3 | |
| 6 | 6 | 18.1 | 18.1 | 18.1 | 15.9 | 9.4 | 12.7 | 23.8 | 8.8 | 16.3 | 25.4 | 15.5 | 20.4 | 14.3 | 9.3 | 11.8 |
| 6 | 7 | 20.2 | 20.2 | 20.2 | 14.1 | 15.1 | 14.6 | 24.9 | 13.9 | 19.4 | 27.2 | 9.6 | 18.4 | 13.7 | 18.0 | 15.8 |
| 6 | 8 | 15.6 | 15.6 | 15.6 | 11.1 | 5.2 | 8.2 | 21.3 | 8.7 | 15.0 | 21.6 | 10.2 | 15.9 | 12.4 | 4.9 | 8.6 |
| 6 | 9 | 24.4 | 24.4 | 24.4 | 14.8 | 7.7 | 11.2 | 18.5 | 20.2 | 19.3 | 17.7 | 19.2 | 18.4 | 12.9 | 10.6 | 11.7 |
| 6 | 10 | 23.4 | 23.4 | 23.4 | 15.9 | 10.5 | 13.2 | 22.9 | 6.1 | 14.5 | 21.8 | 6.9 | 14.4 | 14.4 | 6.7 | 10.5 |
| avg | | | 20.3 | | | 12.0 | | | 16.9 | | | 17.5 | | | 11.7 | |

For a specific quantum program combination, the algorithm computes the EPST violation with the independent EPST and the co-located EPST. If EPST violation is less than the threshold $\varepsilon$ and the number of co-located quantum programs does not exceed the maximum number of co-location, they can be co-located on the chip. In order to ensure the efficiency of the algorithm and the fairness of scheduling, the algorithm only search the first $N$ tasks in the queue. More details are shown in Algorithm 4.

## V. EVALUATION

### A. Methodology

*1) Evaluation metrics:* The following metrics are used for evaluation.

**Probability of a Successful Trial (PST).** PST is used to evaluate the fidelity of the quantum program [7,28,29]. PST is defined as the fraction of trails that produce a correct result. We compile each benchmark and run it on the target quantum chip for 8024 trials.

**Post-compilation gates number.** We use the number of post-compilation quantum gates, **especially** CNOT gates, to show the algorithm's ability to reduce the number of SWAPs when compiling multiple quantum programs.

**Trial Reduction Factor (TRF).** TRF is used to evaluate the improvement of the throughput brought by multi-programming policies [7]. TRF is defined as the ratio of trails needed when programs are executed independently to the trails when multi-programming is enabled.

*2) Benchmarks:* We employ benchmarks used in previous works, including QSAM-Bench [6], RevLib [30] and examples in Alwin Zulehner's work [31]. The benchmarks used for fidelity evaluation are listed in table I, including micro sized benchmarks with several CNOTs and small sized benchmarks with tens of CNOTs.

*3) Quantum chips:* We evaluate our work on IBMQ16 and IBMQ20 platforms. Their architectures are shown in Figure 2. IBMQ16 is publicly available, but IBMQ20 is not. We construct a simulated IBMQ20 quantum chip for the evaluation of post-compilation gates reduction. The gate error rates, T1 relaxation times, and T2 dephasing times of the chip are generated by a uniform distribution random model based on IBMQ16 calibration data.

*4) Baseline:* We compare our techniques, including CDAP and X-SWAP scheme, with following policies.

**Independent execution.** It compiles and executes workloads separately with SABRE [17]. These cases are the most reliable ones without the interference caused by multi-programming.

**Multi-programming baseline.** It uses the policy proposed in [7], which generates initial mapping for a workload with FRP strategy and generates mapping transition with the enhanced noise-aware SABRE strategy.

We show the breakdown of our approach, i.e., CDAP-only and X-SWAP-only, separately. And, we also show the effectiveness of our approach that enables both CDAP and X-SWAP at the same time.

### B. Evaluation results

*1) Evaluation of fidelity:* We use micro/small sized benchmarks for fidelity evaluation. We show PST for combinations include two workloads in Table II. The experiments are all done in a calibration cycle of IBMQ16, which means the calibration data are the same. The combination of two workloads can double the throughput of the quantum computer. However, multi-programming in quantum computers may reduce the reliability due to resource conflicts and crosstalk. In most cases, the average PST of quantum programs executed with multi-programming techniques is lower than that in independent execution cases. In contrast, our approach still outperforms the multi-programming baseline by 12%. The average PST of independent execution, multi-programming baseline and our approach for micro-sized benchmark combinations are 69.9%, 43.7% and 62.8%, respectively. For small-sized benchmark combinations, they are 20.3%, 12.0% and 16.9%. The average PST of our approach decreases by 5.25% on average compared with independent execution cases, which is better than multi-programming baseline's 17.25% decrease, indicating that our approach incurs less fidelity reduction.

TABLE III. NUMBER OF POST-COMPILATION GATES COMPILED WITH DIFFERENT POLICIES ON TWO PLATFORMS.

| Workloads | | IBMQ16 | | | | | IBMQ20 | | | | |
| --- | --- | --- | --- | --- | --- | --- | --- | --- | --- | --- | --- |
| ID1 | ID2 | sum | Baseline | CDAP+X-SWAP | CDAP-only | X-SWAP-only | sum | Baseline | CDAP+X-SWAP | CDAP-only | X-SWAP-only |
| 1 | 1 | 4 | 8 | 4 | 4 | 8 | 4 | 5 | 4 | 4 | 7 |
| 1 | 2 | 5 | 9 | 8 | 8 | 9 | 5 | 8 | 5 | 8 | 8 |
| 1 | 3 | 12 | 25 | 19 | 19 | 25 | 9 | 12 | 10 | 10 | 12 |
| 1 | 4 | 11 | 20 | 9 | 9 | 20 | 8 | 11 | 9 | 9 | 11 |
| 1 | 5 | 13 | 24 | 12 | 12 | 24 | 10 | 10 | 12 | 12 | 12 |
| 6 | 6 | 66 | 72 | 66 | 66 | 72 | 34 | 53 | 53 | 66 | 37 |
| 6 | 7 | 55 | 56 | 55 | 58 | 61 | 31 | 50 | 50 | 50 | 34 |
| 6 | 8 | 69 | 68 | 71 | 73 | 68 | 36 | 52 | 39 | 39 | 36 |
| 6 | 9 | 71 | 81 | 69 | 76 | 81 | 37 | 64 | 50 | 50 | 50 |
| 6 | 10 | 87 | 89 | 83 | 83 | 89 | 39 | 58 | 58 | 58 | 42 |

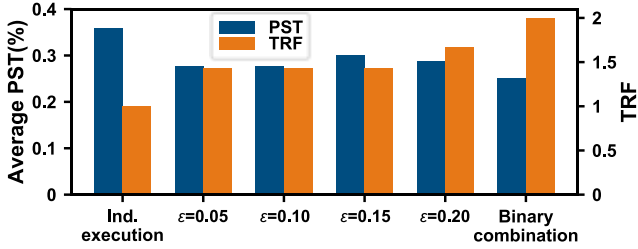

Fig. 13. Performance of the compilation task scheduling framework.

The main benefit brought by our approach is from CDAP strategy. CDAP-only strategy helps to improve the fidelity by providing a better initial mapping. As shown in table II, in workload mix 1 and 4 (i.e., bv_n3 and toffoli_3), CDAP-only strategy provides better initial mapping, improving the fidelity significantly by 44.3% over multi-programming baseline. A good initial mapping makes the fidelity in multi-programming close to or even exceed that in independent execution cases. The advantages of CDAP over multi-programming baseline originate from two aspects: (1) Gates performed on a reliable region have lower error rates. (2) A better initial allocation reduces mapping transition SWAP cost. On average, CDAP-only strategy reduces the impact of multi-programming on fidelity and outperforms baseline by 12.3%.

The X-SWAP-only strategy does not show advantages in fidelity improvement due to the following reasons. (1) In small-sized quantum programs, the initial mapping has a major impact on reliability. The initial mapping of X-SWAP-only cases is generated by the FRP policy, which is the same as the multi-programming baseline. (2) Few SWAPs are needed, and thus X-SWAP scheme becomes useless in the cases where small-sized quantum programs are mapped on a chip with lots of qubits. (3) The allocation of the quantum programs may not be adjacent, and inter-program SWAPs cannot be performed.

*2) Evaluation of post-compilation gates:* We compare the number of post-compilation gates of workload combinations compiled with different strategies on IBMQ16 and IBMQ20 platforms. IBMQ16 and IBMQ20 have different architectures, making the strategy have various benefits on different platforms. The comparison results are in Table III.

Multi-programming on quantum computers may cause resource conflicts between co-located quantum programs, which introduces redundant SWAP operations. Thus, the number of CNOT gates in the programs compiled with multi-programming techniques are greater than the sum of CNOT gates in individually compiled circuits in most cases. Our approach, including CDAP and X-SWAP, helps to reduce the SWAP costs.

On the IBMQ16 platform, CDAP helps to reduce SWAP costs by providing a better initial mapping. Multi-programming baseline and X-SWAP only strategies perform similarly on IBMQ16. This is because they share the same initial mapping generation technique, and the initial mapping counts more on a quantum chip with simple architecture like IBMQ16. The two policies employing CDAP performs similarly for the same reason. Compared with baseline, the post-compilation gates of CDAP-only strategy can reduce by 4.4. CDAP + X-SWAP outperforms baseline by 6.6 in gates reduction. The reduction mainly benefits from an excellent initial allocation, which requires fewer SWAPs to satisfy the constraints of the CNOTs. CDAP helps to find a unique initial mapping, which not only improves the overall fidelity but also helps to reduce the SWAP costs.

On IBMQ20, X-SWAP scheme helps to reduce SWAP costs when processing large-sized benchmark combinations. Compared with baseline, X-SWAP does not show advantages in handling micro-sized benchmark combinations. But, it shows a significant advantage in small-sized benchmark combinations with an average post-compilation gates reduction of 15.6. X-SWAP scheme potentially performs better in the enlarged SWAP search space, which can eliminate the adverse effects of resource conflicts by providing more efficient SWAPs. On average, our approaches helps to save 12.4% post-compilation gates on IBMQ16 and 9.8% post-compilation gates on IBMQ20.

*3) Evaluation of task scheduling:* We constructed a task queue with each of the quantum programs in Table I. The task scheduling framework is evaluated with the queue. In Figure 13, we show the average PST and TRF of the workloads across different EPST violation thresholds. The performance under independent execution cases and random binary combination cases are also shown for comparison.

Independent execution cases can support an average PST of 35.9% with a TRF of 1. In contrast, random binary combination cases have an average PST of 25.1% and a TRF of 2. The PST of our approach is 28.5% on average, which incurs an average loss in fidelity of 7.4%. When $\varepsilon$ equals to 0.15, the average PST reaches a maximum value of 30.1%, and the TRF is 1.43.

The scheduling framework performs best when $\varepsilon$ equals 0.15, which incurs a minimum fidelity reduction of 5.8% and improves the throughput of the quantum computer by 43%. The scheduler trades off between throughput and fidelity, which not only improves the throughput but also incurs less fidelity reduction.

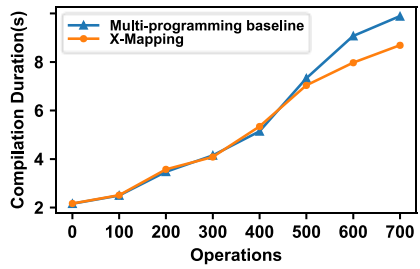

Fig. 14. Compilation duration comparation on average.

*4) Evaluation of compilation time:* Figure 14 compares the average compilation duration of our approach with multi-programming baseline. The experiments are done on a laptop with an Intel ® i7-8550U @1.8GHz core and 8GB memory. We selected combinations of quantum programs with diverse total numbers of gate operations. The workload combinations are compiled with these two strategies and the compilation time are recorded. The overhead of the combined policy is similar to the baseline in small sized benchmarks but shows advantage on larger sized benchmarks with more than 500 operations. Our policy outperforms previous solution by 1.194s on average when handling quantum program combinations with operations up to 700.

## VI. CONCLUSION

Now, diverse quantum programs are developed for new quantum computers. Meanwhile, a quantum chip can have more than tens of qubits than previous days. All of these trends make co-locating multiple quantum programs on a specific chip reasonable. Our work presents a new qubits mapping policy, which takes advantage of the characteristics of multi-programming to improve the fidelity and resource utilization in multi-programming cases. Our approach outperforms the state-of-the-art multi-programming policy by improving the fidelity and reducing the SWAP overheads. As multi-programming is gaining importance, we hope our efforts could be helpful to future researchers in the community.